# Perspectives on spintronic diodes

G. Finocchio[1*], R. Tomasello[2], B. Fang[3], A. Giordano[1], V. Puliafito[4], M. Carpentieri[5], Z. Zeng[6]

[1]Department of Mathematical and Computer Sciences, Physical Sciences and Earth Sciences, University di Messina, I-98166, Messina, Italy

[2]Institute of Applied and Computational Mathematics, FORTH, GR-70013, Heraklion, Crete, Greece

[3] Physical Science and Engineering Division, King Abdullah University of Science and Technology, Thuwal 23955-6900, Saudi Arabia

[4]Department of Engineering, University of Messina, I-98166, Messina, Italy

[5]Department of Electrical and Information Engineering, Polytechnic of Bari, I-70125, Bari, Italy

[6]Key Laboratory of Multifunctional Nanomaterials and Smart Systems, Suzhou Institute of Nano-tech and Nano-bionics, Chinese Academy of Sciences, Ruoshui Road 398, Suzhou 215123, PR China

*Corresponding author: gfinocchio@unime.it

**Abstract**

Spintronic diodes are emerging as disruptive candidates for impacting several technological applications ranging from the Internet of Things to Artificial Intelligence. In this letter, an overview of the recent achievements on spintronic diodes is briefly presented, underling the major breakthroughs that have led these devices to have the largest sensitivity measured up to date for a diode. For each class of spintronic diodes (passive, active, resonant, non-resonant), we indicate the remaining developments to improve the performances as well as the future directions. We also dedicate the last part of this perspective to new ideas for developing spintronic diodes in multiphysics systems by combining 2-dimensional materials and antiferromagnets.

**Overview of the topic**

This perspective will present a brief overview of recent developments on resonant and non-resonant spin-torque diodes (STDs) considering unbiased (passive) and biased (active) configurations and working frequency (GHz or THz). We will focus on open questions, remaining challenges, and future directions for this disruptive technology. The birth of spintronic diodes dates back in 2005[1] with the discovery of the STD effect in MTJs, i.e. a spin-polarized microwave current is converted into a rectified voltage. The interest in STDs has grown for two main reasons: (i) the nanoscale size - they are the smallest rectifiers known in nature -, and (*ii*) CMOS-compatibility, - they can be realized with the same materials used for spin-transfer-torque MRAMs. The performance of an STD can be characterized by different metrics such as sensitivity (defined as the ratio between the rectified output and the input microwave power) that is fundamental for detectors, output resistance, noise equivalent power (NEP), conversion efficiency (defined as the ratio between the dc delivered power over the input microwave power) that is important for energy harvesters. Other crucial aspects are the working temperature, and the requirement of an external bias field. For a detailed review on the history, physical aspects, and working principles of STDs, we suggest Ref. [2].

**STDs working in passive regime.**

*Resonant response.* The resonance response of unbiased STDs is centered near the ferromagnetic resonance frequency (FMR) and, for low input power $P_{in}$, can be described within a linear theory. For example, in an MTJ a microwave spin-polarized current $i_{ac} = I_{ac} \sin(2\pi f_{AC} t + \varphi_I)$, where $I_{ac}$, $f_{AC}$, and $\varphi_I$ are the amplitude, frequency and phase of the current, excites a magnetoresistive signal at the same frequency $r = \Delta R_S \sin(2\pi f_{AC} t + \varphi_R)$ with $\Delta R_S$, and $\varphi_R$ being its amplitude and phase, respectively. The corresponding output voltage is characterized by a dc component $V_{dc} = 0.5 I_{ac} \Delta R_S \cos(\varphi_S)$, where $\varphi_S = \varphi_I - \varphi_R$ is the phase difference between the ac current and the oscillating resistance. There are two main research breakthroughs in material development identified as milestones to realize high-performance STDs: the discovery of the interfacial perpendicular anisotropy (IPA)[3,4], and the voltage-controlled magnetic anisotropy (VCMA).[5,6] The former has been used to design hybrid MTJs with an in-plane (IP) polarizer and an out-of-plane (OOP) free layer,[7,8] while the latter can be used to originate a time-variable magnetic anisotropy which can increase the magnetization precession angle driven by the spin-polarized current $i_{ac}$. By combining both contributions, Zhu et al.[6] demonstrated a significant improvement in the STD performances, which nowadays exhibit input power below nW,[7,9] output resistance smaller than kOhm,[7,10] sensitivity approaching 1000 V/W,[7] and NEP of the same order of Schottky diodes.[7,9] Additionally, for $P_{in}$ approaching nW, resonant unbiased STDs exhibit a conversion efficiency larger than state-of-the art Schottky diodes.[8] The performances of those devices have still room to be improved in different directions, such as reducing the impedance mismatch with smaller Resistance/Area values, and increasing the coefficient of VCMA effect. Other routes concern the use of MTJs with three terminals[11–13] to take advantage of the spin-orbit torque (SOT) and/or bringing topology in the game by using MTJs with non-uniform ground states, such as skyrmions.[14]

The first measurements of non-linear resonant response in unbiased STDs were performed by Sankey et al.[15] who demonstrated, at sufficiently small $i_{ac}$, a transition from linear response to a nonlinear one where the FMR frequency is shifted to higher value and the rectification curve as a function of the microwave frequency becomes asymmetric (Fig. 3(a) in Ref. [15]). This behavior recalls the resonant response of Duffing oscillators.[16]

*Non-resonant response.* Unbiased STDs can be also characterized by a non-resonant response. Up to date, two main solutions are reported: (*i*) broadband response,[8] where a flat rectified output is measured over a wide range of input frequencies; (*ii*) low frequency tail response,[17–19] where the rectified output increases in magnitude as the frequency decreases. Usually, the non-resonant response occurs at frequencies smaller than the ferromagnetic resonance. The realization of non-resonant STDs followed the theoretical predictions by Prokopenko et al.[20] who showed that, in an IP MTJ biased with a sufficiently large OOP field to stabilize a tilted equilibrium configuration, large OOP precessions are exited for input power above a certain threshold.

Experimentally, the key ingredient to obtain a non-resonant response at zero external field (unlike the predictions in Ref. [20]) is the use of hybrid MTJs where the free layer anisotropy field is engineered to work near the compensation point where the IPA is of the same order as the OOP component of the demagnetizing field, and, hence, the second-order anisotropy plays a crucial role in the dynamical properties of the MTJ.[8,19] In the case where the free layer easy-axis is oblique,[8] as the input power increases, a broadband rectification was observed (instead of a non-linear resonance[15]) at sub-gigahertz frequency with a bandwidth of 1.2 GHz.[8] This non-resonant STD was successful for battery-less powering of a black phosphorous nanodevice, thus highlighting a new application of STDs as electromagnetic energy harvesters. Similarly to the theoretical predictions of Ref. [20] the broadband detection occurs for $P_{in}$ larger than a critical value. A broadband response with a nonlinear frequency dependent amplitude can be also driven by an external field applied along the IP hard-axis of the MTJ.[10]

Despite these first proof of concepts, the future challenges for broadband STDs are the increase of the rectification bandwidth in order to collect energy from more signals of environmental electromagnetic background at different frequencies (VSM, broadcast frequencies, etc.), the reduction of the critical power needed to operate at smaller input microwave power, as well as the improvement of the conversion efficiency. Therefore, different device concepts, such as three-terminal device[11–13] with the additional contribution of SOTs, could be exploited to improve the overall performances.

The IPA can be also designed to promote a magnetization IP easy-axis, hence allowing for MHz rectification with STDs. Algarín et al. designed a superparamagnetic MTJ where, thanks to the thermally-activated adiabatic stochastic resonance, a high rectification voltage was measured at a few MHz with a sensitivity close to 100 mV/mW. Tarequzzaman et al.[17] showed a low frequency tail below MHz with a cut-off frequency at about 100 MHz. Tomasello et al.[19] found a low frequency tail due to the excitation of a large amplitude magnetization oscillation with a frequency-dependent phase shift between oscillating current and magnetoresistance. A future direction for non-resonant STD can be the use of ferromagnets with sub-MHz dynamics to develop new device concepts coupling magnetization dynamics with linear and nonlinear mechanical resonances.[21,22]

**STDs working in active regime.**

It has been shown that a bias dc electrical current $I_{dc}$ gives rise to fascinating non-linear effects which can significantly enhance the STD performance, overcoming the thermodynamics limit of the Schottky diodes.[7,9] Within a simple model, the $V_{dc}$ in active STDs is given by two contributions $V_{dc} = 0.5 I_{ac} \Delta R_S \cos(\varphi_S) + I_{dc} \Delta R_{dc}(I_{ac})$, where $\Delta R_{dc}(I_{ac}) = R_{dc}(I_{ac}) - R_{dc}(0)$ is the difference between the average resistance in the presence $R_{dc}(I_{ac})$ and in the absence $R_{dc}(0)$ of $I_{ac}$. The latter term is also frequency-dependent.

Cheng et al.[23] developed a superparamagnetic IP MTJ that, when biased with a proper field and current, has an energy landscape of the free layer magnetization with three minima, two static states and a limit cycle (self-oscillation). The thermal fluctuations drive jumps among the three states with a specific rate which changes as the microwave input is applied. This induces a change in the average resistance, thus enhancing the overall sensitivity of the STD. This solution, which takes advantage of the nonadiabatic stochastic resonance, has achieved a sensitivity larger than 25kV/W, but below room temperature and with the need of an external field.

To obtain large magnetization precession at room temperature, Miwa et al.[9] designed an IP MTJ biased with a field with a certain amplitude and angle to stabilize a tilted equilibrium magnetization configuration $\theta_0$ with respect to the sample plane, and with a current $I_{dc}$ to reduce the effective damping but still working in sub-critical regime (the current does not drive self-oscillations of the magnetization). When the microwave input is applied, the oscillation axis changes direction for frequencies approaching the ferromagnetic resonance, hence changing the average resistance. This solution works at room temperature exhibiting a sensitivity approaching 11 kV/W.

A breakthrough in terms of sensitivity was provided Fang et al.[7]. They have designed a hybrid MTJ (IP polarizer and OOP free layer) to excite a self-oscillations of the free layer magnetization with low dc current.[24] In presence of ac current, the injection locking mechanism is achieved. Because both microwave current and magnetoresistance oscillate at the same frequency, a detection voltage is then observed. However, by fixing the working point of the oscillator near a strongly nonlinear region, where a jump between different oscillation modes occurs, a large sensitivity (> 75kV/W) was observed. This was again possible thanks to the additional contribution originated from the change of the average resistance. Besides the large sensitivity (>210 kV/W) reached for optimized MTJs,[25] the zero-field operation is a significant advantage of STDs based on injection locking over other solutions.

Another strategy used to design high-sensitive STDs is based on MTJs having, at equilibrium, a non-uniform magnetization configuration of the free layer, i.e. a vortex state.[26] The ac current excites the gyrotropic motion of the vortex core in the MHz range, and, as $I_{ac}$ increases, larger orbits are obtained, until the vortex core is expelled from the sample. The core expulsion in presence of a dc current generates an abrupt increase of the detection voltage because of the large variation of the average resistance. The role of dc current is to support the mechanism of vortex core expulsion. Vortex-based STDs have been optimized to achieve a sensitivity of 80 kV/W,[27] however this solution needs an external bias field. Very recently, a sub-GHz high-sensitive STD with a responsivity of $10^6$ V/W under a dc current of -2.6mA and a power of -55 dBm has been realized.[28] The high sensitivity is attributed to nonlinear diode voltage generated by heat-induced spin torque in the presence of a dc current -driven auto-oscillation. To take into account this additional contribution, a dc current given by this thermal effect $I_{th}$ has to be added to the calculation of the rectified volage $V_{dc} = 0.5 I_{ac} \Delta R_S \cos(\varphi_S) + (I_{dc} + I_{th}) \Delta R_{dc}(I_{ac}, I_{th})$. It has to be highlighted that also the change in the dc resistance is dependent on $I_{th}$.

**Roadmap for active STD applications**

It is our opinion that STDs working in active regime are ready to be launched in the market and used massively in a wide range of applications. As a future perspective, additional improvement of active STDs performance and working frequency can be achieved exploiting the large spin-to-charge conversion in heat-driven MTJ, as recently demonstrated for spintronic microwave amplifiers,[29] and already proved in a sub-GHz solution for STDs.[28]

Fig. 1 summarizes the three main applications envisioned for future developments. The first one (Figs. 1(a) and (b)) concerns the use of active STDs as binary frequency shift keying (FSK) demodulators.[30] The FSK modulation, where the digital bits are coded as two microwave signals with different frequencies (see Fig. 1(a), has the advantages to be nearly error-free, and to be implemented with a low energy consumption. These are key requirements for the use in Intern of Things applications. The demodulation scheme can be realized either by two STDs having a proper resonant frequency or by only one STD with a dual-band resonant response,[10] with the benefit to have a more compact solution (see Fig. 1(b) for an example of detection curve).[10] In the dual-band STD, the input signal excites one of the two modes corresponding to the two different frequencies of the STD and, since the rectified voltage has an opposite sign within the two bands, it is possible to reconstruct the sequence of bits.[30]

The recent proof-of-concept of successful communication between a spin-torque nano-oscillator (STNO) and a passive vortex-based resonant STD (Fig. 1(c))[31] has opened the path for the realization of a spintronic transceiver. The STNO generates microwave signals compatible with the detection band of the STD. The microwave signal generated by the STNO is then amplified and converted into a wireless signal. The latter is transmitted and received in the other node where it is converted into a microwave current and rectified by the STD. Fig. 1(d) shows an example of rectification curve of an active STD[7] where, for an input power of 1μW, the rectification voltage reaches values larger than 15mV. The active STD should be used to replace the passive one of Ref. [31], to achieve higher detection sensitivity.[7,32] In general, the advantage of the spintronic transceiver is the compactness, which ensures on-chip integration with a small area occupancy.

The third application envisaged is within the field of neuromorphic computing, where Spintronics and, more specifically, STDs can impact the hardware implementation of neurons and synapses.[34,35] Cai et al.[36] demonstrated that an active STD can be used as a neuron having an activation function with features from both ReLU and soft-plus. It was also proved that a STD based artificial neural network (ANN) can successfully recognize the handwritten digits in the MNIST database with an accuracy of 95.3% with only one hidden layer. Leroux et al.[33] proposed the use of passive resonant

STDs as tunable synapsis to connect the layers of radio-frequency neurons in ANNs. The key aspect is that the synaptic weight depends on the difference between the frequency of the input signal and the resonance frequency of the STD. It was theoretically proved the feasibility to perform a Multiply-And-Accumulate (MAC) operation, which is at the basis of the synapsis functioning, and a ANNs made of 10 chains of 64 STDs (Fig. 1(e)) was tested on a dataset called "Digits" of handwritten digits, with a resulting accuracy of 99.96%. In addition, Fig. 1(f) highlights that even nonlinear STDs with proper nonlinear coefficients can be used for this purpose, due to the coincidence of the output voltage with the linear STDs. To impact seriously in neuromorphic computing, the device-to-device variations of tunneling magnetoresistance and magnetic parameters of STDs with respect to the nominal values have to be reduced in order to have a robust approach for performing MAC operation.

**Overview of spintronic THz detectors and challenges**

A new direction for STDs is in field of Antiferromagnetic Spintronics,[37,38] which promises to move the working frequencies from microwave to THz, with the additional advantages of robustness against external field, scalability because of the absence of dipolar fields, and tunability with current.[39–41] This range of frequencies, in fact, after decades of little interest, has turned out very appropriate for lots of applications ranging from imaging, medical and security to Information and Communications Technology. While antiferromagnetic switching at THz has been already demonstrated experimentally,[42] most device concepts for THz Spintronics, such as detectors, still remain on the paper.[43–45]

Most of the proposals for antiferromagnetic STDs are based on multiterminal devices where the SOT is the main mechanism driving the dynamics. Khymyn et al.[43] predicted passive STD with a sensitivity of $10^2$-$10^3$ V/W, comparable with the Schottky diodes, in nanoscale materials with bi-axial anisotropy. Active narrow band THz detectors[46] can be realized by taking advantage of the injection locking between a persistent magnetization dynamics excited by large enough currents, and a THz current, with the perspective to strongly enhance the sensitivity, similarly to what occurs in ferromagnetic detectors,[7,32] and to define a narrow detection band with a quality factor of about 400, as shown in Fig. 2(a). Current-tunable THz detectors[45] can be obtained by biasing the device in a sub-critical regime (Fig. 2(b)). The theoretical predictions of sensitivity of the order of $10^3$ V/W, and electrical tunability (unique property) make THz STDs candidates for challenging graphene-based solutions which represent a new standard in the field of THz detectors.[47–49]

The main challenges for the experimental proof-of-concept of THz spintronic detectors can be summarized as (i) stabilization of single domain thin films of antiferromagnetic materials interfaced with heavy metal or exhibiting intrinsic SOT, (ii) control of magnetoelastic energy, and (iii) improvement of the read-out mechanism. To achieve (i), material development and nanofabrication challenges have to be faced to scale device size from micrometer[50] to nanometer dimensions. Point (ii) is a dominant factor in determining the antiferromagnetic domains and can drive local variation of magnetic parameters with the creation of magnetic grains with size of the order of tens of nanometers. A direction to tackle (iii) can be a generalization of what proposed by Safin et al.[45] and then involving inverse spin-Hall effect, and spin-pumping due to AFMs and heavy metals, such as Pt (see Fig. 2(c)). An alternative solution can be based on a three-terminal AFM tunnel junction.[51,52]

**Perspective on multiphysics detectors**

As a final remark, we wish to highlight that integration between 2-dimensional materials[53,54] and spintronic devices seems feasible and very intriguing for the next-years roadmap. Those hybrid systems can take advantage of multiphysics coupling to improve, on one hand, the detection performance, flexibility, and scalability of THz detectors and, on the other hand, to drive new fundamental knowledges.

For example, devices combining graphene and AFM could exploit magnetoelastic energy to control the ground state of AFMs (see Fig. 2(d)). In those devices, it will be fascinating look at the excitation of collective MHz/GHz modes coupling plasmonic resonance in graphene and magnetic dynamics in AFM for a new generation of THz detector having advantage of both materials.

**Acknowledgements**

This work is supported under Grant No. 2019-1-U.0. ("Diodi spintronici rad-hard ad elevate sensitività - DIOSPIN") funded by the Italian Space Agency (ASI) within the call "Nuove idee per la componentistica spaziale del futuro" and the Executive Programme of Scientific and Technological Cooperation Between Italy and China (2016YFE0104100). R.T. and G.F. thank the project "ThunderSKY" funded by the Hellenic Foundation for Research and Innovation and the General Secretariat for Research and Technology, under Grant No. 871. This work was partially supported by the National Science Foundation of China (No. 11804370).

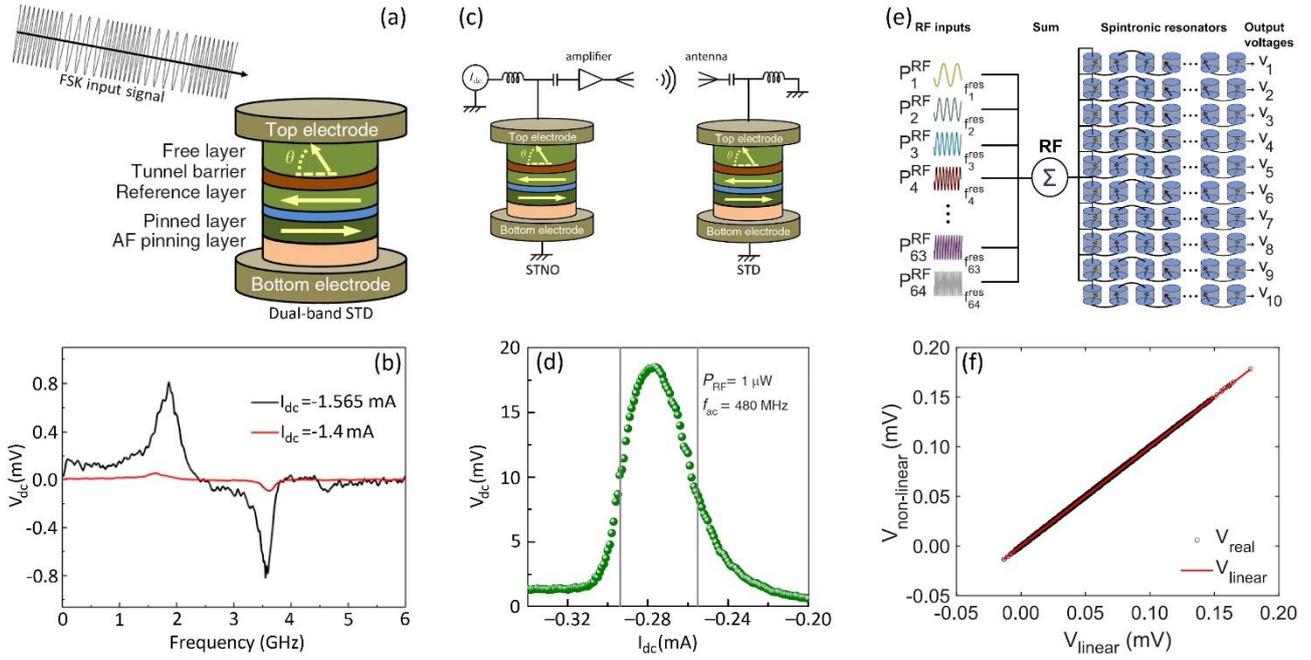

Figure 1: (a) Sketch of a dual-band detection of FSK-modulated input signal. B. Fang et al. Phys. Rev. Appl. 11, 014022 (2019);[7] licensed under a Creative Commons Attribution (CC BY) license. (b) Resonant response of a dual-band STD. Reproduced from L. Zhang et al. Appl. Phys. Lett. 117, 072409 (2020),[10] with the permission of AIP Publishing. (c) Sketch of a spintronic transceiver, similarly to Ref. [31]. (d) Resonant response of an active STD with injection locking. B. Fang et al. Phys. Rev. Appl. 11, 014022 (2019);[7] licensed under a Creative Commons Attribution (CC BY) license. (e) Sketch of ANN based on STD tunable synapsis. (f) Comparison of the output voltage of the simulated tunable synapsis with linear and nonlinear STDs. (e) and (f) N. Leroux et al., Arxiv:2011.07885 (2020);[33] licensed under a Creative Commons Attribution (CC BY) license.

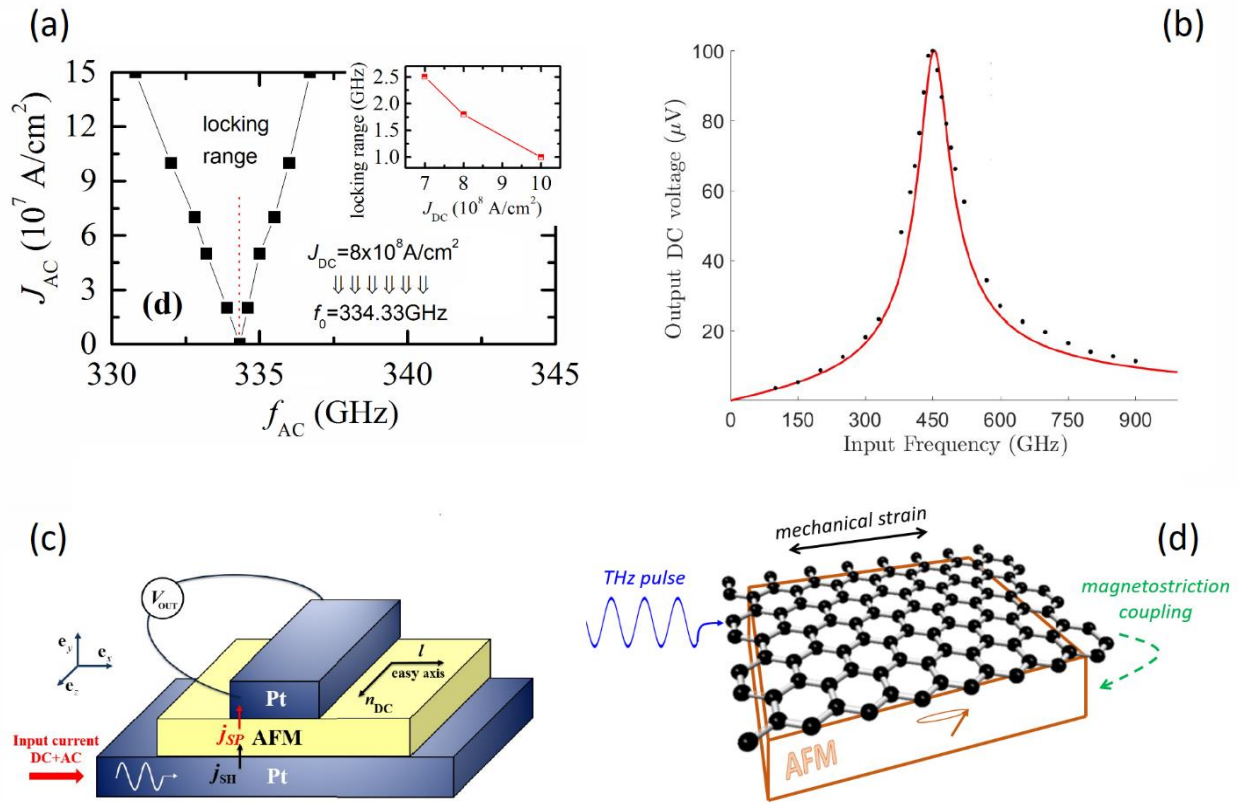

Figure 2: (a) Injection locking band of an AFM STNO, where the inset show the injection locking range as a function of the dc current. Reproduced from V. Puliafito et al., IEEE Trans. Magn. 57, 4100106 (2021),[46] with the permission of IEEE. (b) Resonant response of an AFM SOT THz detector. (c) Sketch of the AFM THz detector based on SOT. (b) and (c) are reproduced from A. Safin et al., Appl. Phys. Lett. 117, 222411 (2020),[45] with the permission of AIP Publishing. (d) Sketch of a bilayer combining a graphene sheet with an AFM layer.